\documentclass[british,conference]{llncs}
\usepackage[latin9]{inputenc}
\usepackage{float}
\usepackage{wrapfig}
\usepackage{amsmath}
\usepackage{amssymb}
\usepackage{graphicx}
\usepackage{hyperref}
 \hypersetup{pdftitle={Lightweight Monte Carlo Verification of Markov Decision Processes with Rewards}}

\makeatletter

\providecommand{\tabularnewline}{\\}

 \usepackage[ruled,vlined]{algorithm2e}
 \usepackage{tikz}

\LinesNumbered
\let\oldnl\nl
\newcommand{\nonl}{\renewcommand{\nl}{\let\nl\oldnl}}

\makeatother

\usepackage{babel}
\begin{document}

\title{Lightweight Monte Carlo Verification of Markov Decision Processes
with Rewards}

\author{Axel Legay, Sean Sedwards and Louis-Marie Traonouez}

\institute{Inria Rennes -- Bretagne Atlantique}
\maketitle
\begin{abstract}
Markov decision processes are useful models of concurrency optimisation
problems, but are often intractable for exhaustive verification methods.
Recent work has introduced lightweight approximative techniques that
sample directly from scheduler space, bringing the prospect of scalable
alternatives to standard numerical model checking algorithms. The
focus so far has been on optimising the probability of a property,
but many problems require quantitative analysis of rewards. In this
work we therefore present lightweight statistical model checking algorithms
to optimise the rewards of Markov decision processes. We consider
the standard definitions of rewards used in model checking, introducing
an auxiliary hypothesis test to accommodate reachability rewards.
We demonstrate the performance of our approach on a number of standard
case studies.
\end{abstract}

\section{Introduction }

Markov decision processes (MDP) describe systems that interleave nondeterministic
actions and probabilistic transitions. Such systems may be seen as
comprising probabilistic subsystems whose transitions depend on the
states of the other subsystems, while the order in which concurrently
enabled transitions execute is nondeterministic. This order, defined
by a scheduler that is typically either history-dependent or memoryless,
may radically affect the system's behaviour. By assigning numerical
rewards or costs to execution traces, MDPs have proven useful in many
real optimisation problems \cite{White1993}. More recently, in the
context of formal verification, logics have been extended to allow
model checkers to consider rewards \cite{KwiatkowskaNormanParker2007}.

In the classic context, rewards are assigned to actions \cite{Bellman1957,Puterman1994,BaierKatoen2008}.
In the context of model checking, rewards are often assigned to states
or transitions between states \cite{KwiatkowskaNormanParker2007}.
In both cases the rewards are summed over the length of a trace and
the expected reward is calculated by averaging the total reward with
respect to the probability of the trace. In this work we focus on
MDPs in the context of model checking, but the mechanism of accumulating
rewards is unimportant to our algorithms and we simply assume that
a total reward is assigned to a finite trace.

\begin{wrapfigure}{o}{0.44\columnwidth}%
\centering
\begin{tikzpicture}[every node/.style={circle,inner sep=1pt}]
\draw
(0,0)node[draw](0){$s_0$}
(-1,-2)node[draw](5){$s_1$}
(1,-2)node[draw](3){$s_2$};
\draw
(0)node[above right=0.5em]{$\models\phi$}
(3)node[rectangle,below=0.8em]{$\models\psi$}
(5)node[rectangle,below=0.8em]{$\models\psi\wedge\phi$};
\draw
(-0.5,-1)node[fill=black](2){}
(0.5,-1)node[fill=black](4){}
(1.3,-0.7)node[fill=black](1){}
(-1.3,-0.7)node[fill=black](6){};
\draw[->]
(0,0.75)--(0)
(4)edgenode[right]{$q$}(3)
(2)edgenode[above left]{$1$}(5)
(4)edgenode[below right]{$p$}(5)
(1)edge[bend right=20]node[above right]{1}(0)
(6)edge[bend left=20]node[above left]{1}(0);
\draw
(0)edgenode[left]{$a_1\,r_1$}(2)
(0)edgenode[right]{$a_2\,r_2$}(4)
(3)edge[bend right=39]node[right]{$a_0\,r_0$}(1)
(5)edge[bend left=39]node[left]{$a_0\,r_0$}(6);
\end{tikzpicture}\caption{Simple MDP with different rewards for history-dependent and memoryless
schedulers.\label{fig:historydependent}}

\end{wrapfigure}
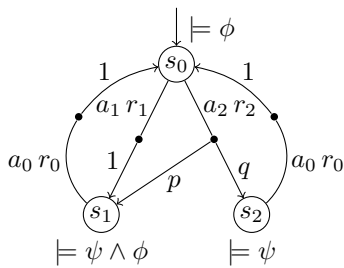

Figure \ref{fig:historydependent} illustrates a simple MDP for which
memoryless and history-dependent schedulers can give different minimum
rewards for logical property $\mathbf{X}(\psi\wedge\mathbf{XG}^{4}\phi)$
(the precise semantics of this logic are defined in Section \ref{sec:preliminaries}).
Execution proceeds by first choosing an action nondeterministically,
to select a distribution of probabilistic transitions, and then by
making a probabilistic choice to select the next state. The property
asserts that on the next step $\psi$ will be true and, on the step
after that, $\phi$ will be remain true for $4$ time steps. In this
example rewards $r_{0},r_{1},r_{2}$ are assigned to actions $a_{0},a_{1},a_{2}$,
respectively. In the initial state ($s_{0}$) both actions $a_{1}$
and $a_{2}$ can lead to traces that satisfy the property, but subsequent
actions taken in state $s_{0}$ must be $a_{1}$. If $r_{1}>r_{2}>0$,
the minimum reward will be achieved by taking action $a_{2}$ in the
initial state and $a_{1}$ whenever the execution visits $s_{0}$
thereafter. To satisfy the property a memoryless scheduler would be
forced to always take action $a_{1}$ in state $s_{0}$ and would
therefore not achieve the minimum possible reward.

Numerical model checking algorithms for probabilistic systems have
complexity related to the number of states of the model and therefore
scale exponentially with the number of interacting variables \cite{BiancoDeAlfaro1995}.
Numerical algorithms to find optimal schedulers in MDPs incur the
additional costs of optimisation \cite{KwiatkowskaNormanParker2007}.
Statistical model checking (SMC) describes a collection of Monte Carlo
techniques that approximate the results of numerical model checking
and aim not to construct an explicit representation of the state space---states
are generated on the fly during simulation. While memory-efficient
(``lightweight'') SMC techniques have been developed to address
the probabilistic model checking problem, until recently \cite{LegaySedwardsTraonouez2014}
it has not been possible to address the nondeterministic problem in
this way. In addition to not incurring the cost of exploring the state
space, a significant advantage of lightweight SMC approaches is that
they may be efficiently distributed on memory-sensitive high performance
parallel computing architectures, such as GPGPU (general purpose computing
on graphics processing units).

SMC makes use of an executable model, so to apply SMC to MDPs the
nondeterminism must be resolved by a scheduler. Since nondeterministic
and probabilistic choices are interleaved in an MDP, memoryless schedulers
are typically of the same order of complexity as the system as a whole
and may be infinite. History-dependent schedulers are exponentially
bigger. The key contribution of \cite{LegaySedwardsTraonouez2014}
is the introduction of techniques to define the behaviour of schedulers
in $\mathcal{O}(1)$ memory, allowing them to be sampled at random
and tested individually. These techniques offer a significant saving
in computation over enumeration, with performance effectively independent
of the size of the sample space and only dependent on the relative
abundance of `near-optimal' \cite{KearnsMansourNg2002} schedulers.

The basic idea of \cite{LegaySedwardsTraonouez2014} is to select
a number of schedulers at random and apply SMC to each of the discrete
time Markov chains they induce. In this way it is possible to estimate
or test hypotheses about the maximum and minimum probabilities of
a property. The simple sampling strategies of \cite{LegaySedwardsTraonouez2014}
have been significantly enhanced by so-called ``smart sampling''
in \cite{D'ArgenioLegaySedwardsTraonouez2015}. Smart sampling makes
gains in performance by refining a candidate set of schedulers and
not wasting simulation budget on those that are obviously sub-optimal.

In this work we present algorithms to find schedulers that approximately
maximise or minimise the expected reward of finite traces in Markov
decision processes. Our algorithms are based on the elemental sampling
strategies given in \cite{LegaySedwardsTraonouez2014} and the smart
sampling techniques of \cite{D'ArgenioLegaySedwardsTraonouez2015}.
The passage from probabilities to rewards is intuitive, but not immediate,
due to the specific way rewards are defined in the context of model
checking \cite{KwiatkowskaNormanParker2007} and to the fact that
the values of rewards are arbitrary. Unlike probabilities, rewards
have no inherent a priori bounds, so the standard statistical techniques
to bound the absolute error do not apply. Our solution is to use a
relative bound, based on a generalisation of the standard Chernoff-Hoeffding
bound used in SMC. A further challenge is that the commonly used `reachability
rewards' \cite{KwiatkowskaNormanParker2007} assume the probability
of a property is known with absolute certainty and, somewhat arbitrarily,
define the reward of properties with probability less than $1$ to
be infinite. With no a priori information about the property, this
definition induces an unknown distribution with potentially infinite
variance. As such, its properties cannot be directly estimated by
sampling. Our solution is to introduce an auxiliary hypothesis test
to assert that the probability of the property is $1$. The estimation
results can then be said to lie within the specified confidence bounds
given that the hypothesis is true, while the confidence of the hypothesis
test can be made arbitrarily high.

We have implemented the algorithms in our statistical model checking
platform, \textsc{Plasma} \cite{JegourelLegaySedwards2012,Boyer-et-al2013}%
\footnote{projects.inria.fr/plasma-lab/reward-estimation\label{fn:plasma}%
}, and have applied them to a number of case studies from the literature.
The results demonstrate that our approach is highly effective on many
models and produces useful bounds when near-optimal schedulers are
rare.

\subsection*{Related Work\label{sec:related}}

The exponential scaling of numerical algorithms, such as policy iteration
and value iteration \cite{Bellman1957,Puterman1994}, has prompted
much work on sampling techniques to approximately optimise discounted
rewards over infinite horizons (see, e.g., \cite{ChangHuFuMarcus2013}
for a survey). Of this work, the Kearns algorithm \cite{KearnsMansourNg2002}
is relevant to the present context because it is lightweight in concept
and used by \cite{LassaignePeyronnet2012} in the context of SMC.
To find the action with the greatest expected reward in the current
state, the algorithm recursively estimates the rewards of successive
states using sampling, until successive estimates differ by less than
some user-defined threshold.

The rewards used in model checking are typically not discounted and
defined over finite horizons \cite{KwiatkowskaNormanParker2007}.
There is much less work on approximative techniques for the finite
horizon problem in the classic literature. Below we summarise several
recent attempts to apply SMC to MDPs \cite{Bogdoll2011,LassaignePeyronnet2012,Henriques-et-al2012,HartmannsTimmer2013,LegaySedwardsTraonouez2014,Brazdil-et-al2014,D'ArgenioLegaySedwardsTraonouez2015},
although none of these address the standard model checking problems
of MDPs with rewards.

In \cite{Bogdoll2011,HartmannsTimmer2013} the authors present algorithms
to remove `spurious' nondeterminism on the fly, so that standard
SMC may be used. This approach is limited to the class of MDPs whose
nondeterminism is not affected by scheduling.

In \cite{Henriques-et-al2012} the authors count the occurrence of
state-actions in simulations, to iteratively improve a probabilistic
scheduler that is assessed using sequential hypothesis testing. If
an example is found it is correct, but the frequency of state-actions
is not in general indicative of global optimality.

In \cite{LassaignePeyronnet2012} the authors use an adaptation of
the Kearns algorithm to find a memoryless scheduler that is near optimal
with respect to a discounted reward scheme. The resulting scheduler
induces a Markov chain whose properties may be verified with standard
SMC. Such properties are only indirectly related to the original MDP.

In \cite{Brazdil-et-al2014} the authors present learning algorithms
to bound the maximum probability of (unbounded) reachability properties.
The algorithms refine upper and lower bounds associated to individual
state-actions, according to the contribution of individual simulations.
The algorithms converge very slowly and may not converge to the global
optimum for similar reasons to those affecting \cite{Henriques-et-al2012}.

The above SMC approaches use data structures whose size scales with
the state space of the MDP. In \cite{LegaySedwardsTraonouez2014}
the authors introduce lightweight techniques to sample directly from
scheduler space using constant memory. The simple sampling strategies
of \cite{LegaySedwardsTraonouez2014} are made more efficient in \cite{D'ArgenioLegaySedwardsTraonouez2015}.
The present work builds on the results of \cite{LegaySedwardsTraonouez2014}
and \cite{D'ArgenioLegaySedwardsTraonouez2015}, which are reviewed
in Section \ref{sec:lightweight}.

\section{Preliminaries\label{sec:preliminaries}}

In this work an MDP comprises a possibly infinite set of states $S$,
a finite set of actions $A$, a finite set of probabilities $Q$ and
a relation $T:S\times A\times S\times Q$, such that $\forall s\in S$
and $\forall a\in A$, $\sum_{\forall s'\in S}T(s,a,s')=r$, where
$r\in\{0,1\}$. The execution of an MDP proceeds by a sequence of
transitions between states, starting from an initial state, inducing
a set of possible traces $\Omega=S^{+}$. Given an MDP in state $s$,
an action $a$ is chosen nondeterministically from the set $\{a'\in A:\sum_{\forall s'\in S}T(s,a',s')=1\}$.
A new state $d\in S$ is then chosen at random with probability $T(s,a,d)$.
We assume that rewards are defined by some function $R:S^{+}\rightarrow\mathbb{Q}$
or $R:A^{+}\rightarrow\mathbb{Q}$ that maps a sequence of states
or a sequence of actions to a total reward. In what follows we abuse
the notation and simply write $R(\omega)$ to mean the total reward
assigned to trace $\omega\in\Omega$ according to an arbitrary reward
scheme.

We later present algorithms to find deterministic schedulers that
approximately maximise or minimise expected rewards for an MDP. A
history-dependent scheduler is a function $\mathfrak{S}:\Omega\rightarrow A$.
A memoryless scheduler is a function $\mathfrak{M}:S\rightarrow A$.
Intuitively, at each state in the course of an execution, a history-dependent
scheduler chooses an action based on the sequence of previous states
and a memoryless scheduler chooses an action based only on the current
state. History-dependent schedulers therefore include memoryless schedulers.
In the context of SMC we consider finite simulation traces of bounded
length, hence $\mathfrak{S}$ and $\mathfrak{M}$ are finite.

In this work we assume properties are defined in a bounded linear
time temporal logic with the following syntax:
\begin{equation}
\phi=\phi\vee\phi\mid\phi\wedge\phi\mid\neg\phi\mid\mathbf{X}\phi\mid\mathbf{F}^{k}\phi\mid\mathbf{G}^{k}\phi\mid\phi\mathbf{U}^{k}\phi\mid\alpha\label{eq:BLTL}
\end{equation}
The symbol $\alpha$ denotes an atomic property that is either \emph{true}
or \emph{false} in a state. Given a trace $\omega\in\Omega$, comprising
states $s_{0}s_{1}\dots$, $\omega^{(i)}$ denotes the trace suffix
$s_{i}s_{i+1}\dots$. The satisfaction relation $\models$ over (\ref{eq:BLTL})
is constructed inductively as follows:
\begin{equation}
\begin{aligned}\begin{split}\omega^{(i)}\models\, & true\\
\omega^{(i)}\models\, & \alpha\iff\alpha\textnormal{ is }\mathit{true}\textnormal{ in state }\omega_{i}\\
\omega^{(i)}\models\, & \neg\varphi\iff\omega^{(i)}\models\varphi\not\in\,\models\\
\omega^{(i)}\models\, & \varphi_{1}\vee\varphi_{2}\iff\omega^{(i)}\models\varphi_{1}\textnormal{ or }\omega^{(i)}\models\varphi_{2}\\
\omega^{(i)}\models\, & \mathbf{X}^{k}\varphi\iff\omega^{(k+i)}\models\varphi\\
\omega^{(i)}\models\, & \varphi_{1}\mathbf{U}^{k}\varphi_{2}\iff\exists j\in\{i,\dots,i+k\}:\omega^{(j)}\models\varphi_{2}\\
 & \wedge(j=i\vee\forall l\in\{i,\dots,j-1\}:\omega^{(l)}\models\varphi_{1})
\end{split}
\end{aligned}
\label{eq:semantics}
\end{equation}
Other elements of the relation are constructed using the equivalences
$\mathit{false}\equiv\neg\mathit{true}$, $\phi\wedge\phi\equiv\neg(\neg\phi\vee\neg\phi)$,
$\mathbf{F}^{k}\phi\equiv\mathit{true}\mathbf{U}^{k}\phi$, $\mathbf{G}^{k}\phi\equiv\neg(\mathit{true}\mathbf{U}^{k}\neg\phi)$.

SMC algorithms typically work by constructing an automaton to decide
the truth of the statement $\omega\models\varphi$, i.e., whether
simulation trace $\omega$ satisfies property $\varphi$. The expected
probability of $\varphi$ is then estimated by $\frac{1}{N}\sum_{i=1}^{N}\mathbf{1}(\omega_{i}\models\varphi)$,
where $\omega_{1},\dots,\omega_{N}$ are $N$ statistically independent
random simulation traces and $\mathbf{1}:\{\mathit{true},\mathit{false}\}\rightarrow\{0,1\}$
is an indicator function that returns $1$ if its argument is \emph{true}
and $0$ otherwise. To bound the estimation error it is common to
use the ``Chernoff'' bound of \cite{Okamoto1958}. The user specifies
an absolute error $\varepsilon$ and a probability $\delta$ to define
the bound $\mathrm{P}(\mid\hat{p}-p\mid\geq\varepsilon)\leq\delta$,
where $p$ and $\hat{p}$ are respectively the true probability and
the estimated probability. The bound is guaranteed if the number of
simulations $N$ satisfies the relation
\begin{equation}
N\geq\left\lceil (\ln2-\ln\delta)/(2\varepsilon^{2})\right\rceil .\label{eq:chernoff}
\end{equation}

\section{Lightweight Verification with Smart Sampling\label{sec:lightweight}}

We recall here the techniques of lightweight verification presented
in \cite{LegaySedwardsTraonouez2014} and \cite{D'ArgenioLegaySedwardsTraonouez2015}.

\subsection{Pseudo-Random Number Generators}

To avoid storing schedulers as explicit mappings, we construct schedulers
on the fly using uniform pseudo-random number generators (PRNG) that
are initialised by a seed and iterated to generate the next pseudo-random
value. Our technique uses two independent PRNGs that respectively
resolve probabilistic and nondeterministic choices. The first is used
in the conventional way to make pseudo-random choices during a simulation
experiment. The second PRNG is used to choose actions such that the
choices are consistent between different simulations in the same experiment.
Given multiple simulation experiments, the further role of the second
PRNG is to range uniformly over all possible sets of choices. The
seed of the second PRNG can be seen as the identifier of a specific
scheduler.

To estimate the probability of a property under a scheduler, we generate
multiple probabilistic simulation traces by fixing the seed of the
PRNG for nondeterministic choices while choosing random seeds for
the PRNG for probabilistic choices. To ensure that we sample from
history-dependent schedulers, we construct a per-step PRNG seed that
is a hash of a large integer representing the sequence of states up
to the present and a specific scheduler identifier \cite{LegaySedwardsTraonouez2014}.

\subsection{Hashing the Trace}

We assume that the state of an MDP is an assignment of values to a
vector of $n$ system variables $v_{i},i\in\{1,\dots,n\}$, with each
$v_{i}$ represented by a number of bits $b_{i}$. The state can thus
be represented by the concatenation of the bits of the system variables,
while a sequence of states (a trace) may be represented by the concatenation
of the bits of all the states. We interpret such a sequence of states
as an integer of $\sum_{i=1}^{n}b_{i}$ bits, denoted $\overline{s}$,
and refer to this as the \emph{trace vector}. A scheduler is denoted
by an integer $\sigma$ of $b_{\sigma}$ bits, which is concatenated
to $\overline{s}$ (denoted $\sigma:\overline{s}$) to uniquely identify
a trace and a scheduler. Our approach is to generate a hash code $h=\mathcal{H}(\sigma:\overline{s})$
and to use $h$ as the seed of a PRNG that resolves the next nondeterministic
choice. In this way we can approximate the scheduler functions $\mathfrak{S}$
and $\mathfrak{M}$: $\mathcal{H}$ maps $\sigma:\overline{s}$ to
a seed that is deterministically dependent on the trace and the scheduler;
the PRNG maps the seed to a value that is uniformly distributed but
also deterministically dependent on the trace and the scheduler. Algorithm
\ref{alg:simulate} implements these ideas as a simulation function
that returns a trace, given a scheduler and bounded temporal property
as input. The uniformity of scheduler selection is demonstrated by
the accuracy of the estimates labelled `uniform prob.' in Fig. \ref{fig:gossip}.

\begin{algorithm}
\KwIn{\\\Indp

$\mathcal{M}$: an MDP with initial state $s_{0}$

$\varphi$: a bounded temporal logic property

$\sigma$: an integer identifying a scheduler

}\KwOut{\\\Indp

$\omega$: a simulation trace

}\BlankLine

Let $\mathcal{U}_{\mathrm{prob}},\mathcal{U}_{\mathrm{nondet}}$ be
uniform PRNGs with respective samples $r_{\mathrm{pr}},r_{\mathrm{\mathrm{nd}}}$

Let $\mathcal{H}$ be a hash function

Let $s$ denote a state, initialised $s\leftarrow s_{0}$

Let $\omega$ denote a trace, initialised $\omega\leftarrow s$

Let $\overline{s}$ be the trace vector, initially empty

Select seed of $\mathcal{U}_{\mathrm{prob}}$ randomly

\While{$\omega\models\varphi$ is not decided}{

$\overline{s}\leftarrow\overline{s}:s$ 

Set seed of $\mathcal{U}_{\mathrm{nondet}}$ to $\mathcal{H}(\sigma:\overline{s})$

Iterate $\mathcal{U}_{\mathrm{nondet}}$ to generate $r_{\mathrm{nd}}$
and use to resolve nondeterministic choice

Iterate $\mathcal{U}_{\mathrm{prob}}$ to generate $r_{\mathrm{pr}}$
and use to resolve probabilistic choice

Set $s$ to the next state

$\omega\leftarrow\omega:s$

}

\caption{Simulate\label{alg:simulate}}
\end{algorithm}

\subsection{Implementation}

To implement our approach we use an efficient hash function that constructs
seeds incrementally using standard precision mathematical operations.
The function is based on modular division \cite[Ch. 6]{Knuth1998},
such that $h=(\sigma:\overline{s})\bmod m$, where $m$ is a large
prime not close to a power of 2 \cite[Ch. 11]{CormenLeiersonRivestStein2009}.
Since $\overline{s}$ is typically very large, we use Horner's method
\cite{Horner1819}\cite[Ch. 4]{Knuth1998} to generate $h$: we set
$h_{0}=\sigma$ and find $h\equiv h_{n}$ ($n$ as above) by iterating
the recurrence relation 
\begin{equation}
h_{i}=(h_{i-1}2^{b_{i}}+v_{i})\bmod m.\label{eq:horner}
\end{equation}

Equation (\ref{eq:horner}) allows us to generate a hash code knowing
only the current state and the hash code from the previous step. When
considering memoryless schedulers we need only know the current state.
Using suitable congruences \cite{LegaySedwardsTraonouez2014}, the
following equation allows (\ref{eq:horner}) to be implemented using
efficient native arithmetic:

\[
(h_{i-1}2^{j})\bmod m=(h_{i-1}2^{j-1})\bmod m+(h_{i-1}2^{j-1})\bmod m
\]

In a typical implementation on current hardware, a hash function and
PRNG may span around $10^{19}$ schedulers. This is usually many orders
of magnitude more than the number of schedulers sampled. There is
no advantage in sampling from a larger set of schedulers until the
number of samples drawn approaches the size of the sample space.

\subsection{Multiple Estimates}

To avoid the cumulative error when choosing a single probability estimate
from a number of alternatives, \cite{LegaySedwardsTraonouez2014}
defines the following Chernoff bound for multiple estimates:

\begin{equation}
N\geq\left\lceil \left(\ln2-\ln\left(1-\sqrt[M]{1-\delta}\right)\right)/(2\varepsilon^{2})\right\rceil .\label{eq:NMChernoff}
\end{equation}

Given $M$ estimates $\{\hat{p}_{1},\dots,\hat{p}_{M}\}$ of corresponding
true probabilities $\{p_{1},\dots$, $p_{M}\}$ each generated with
$N$ samples, (\ref{eq:NMChernoff}) asserts that for any estimate
$\hat{p}_{i}$, in particular the minimum or maximum, $\mathrm{P}(\mid\hat{p}_{i}-p_{i}\mid\geq\varepsilon)\leq\delta$.
Note that when $M=1$, (\ref{eq:NMChernoff}) degenerates to (\ref{eq:chernoff}).

\subsection{Smart Sampling\label{sec:smart}}

The elemental sampling strategies presented in \cite{LegaySedwardsTraonouez2014}
have the disadvantage that they allocate equal simulation budget to
all schedulers, regardless of their merit. Intuitively, the performance
of a scheduler may become apparent, if not certain, long before all
its simulation budget has been used. The idea of smart sampling is
to not waste budget on schedulers that are clearly not optimal, to
thus maximise the probability of finding an optimal scheduler with
a finite simulation budget. We recall here the basic notions of smart
sampling introduced in \cite{D'ArgenioLegaySedwardsTraonouez2015}. 

In general, the problem of finding optimal schedulers using sampling
has two independent components: the rarity of near-optimal schedulers
(denoted $p_{g}$) and the average probability of the property under
near-optimal schedulers (denoted $p_{\overline{g}}$). A near-optimal
scheduler is one whose reward or probability (depending on the context)
is within some $\varepsilon$ of the optimal value. If we select $M$
schedulers at random, using the techniques presented above, and verify
each with $N$ simulations, the expected number of traces that satisfy
the property using a near-optimal scheduler is thus $Mp_{g}Np_{\overline{g}}$
. The probability of seeing a trace that satisfies the property using
a near-optimal scheduler is the cumulative probability 
\begin{equation}
(1-(1-p_{g})^{M})(1-(1-p_{\overline{g}})^{N}).\label{eq:probgoodsched}
\end{equation}

To maximise the chance of seeing a good scheduler with a simulation
budget of $N_{\max}=NM$, $N$ and $M$ should be chosen to maximise
(\ref{eq:probgoodsched}). Then, following a sampling experiment using
these values, any scheduler that produces at least one trace that
satisfies $\varphi$ becomes a candidate for further investigation.
Since the values of of $p_{g}$ and $p_{\overline{g}}$ are unknown
a priori, it is necessary to perform an initial uninformed sampling
experiment to estimate them, setting $N=M=\lceil\sqrt{N_{\max}}\rceil$.
The results can be used to numerically optimise (\ref{eq:probgoodsched}),
however an effective heuristic is to set $N=\lceil1/\hat{p}_{\overline{g}}\rceil$,
where $\hat{p}_{\overline{g}}$ is the maximum observed estimate (or
minimum non-zero estimate in the case of finding minimising schedulers).

The best scheduler is found by iteratively refining the candidate
set. At each iteration the per-iteration simulation budget ($N_{\max}$)
is divided between the remaining candidates, simulations are performed
and the average reward for each scheduler is estimated. Schedulers
whose estimates fall into the ``worst'' quantile (lower or upper
half, depending on context) are discarded. Refinement continues until
estimates are known with specified confidence, according to (\ref{eq:NMChernoff}).
With a per-iteration budget satisfying (\ref{eq:chernoff}), the algorithm
is guaranteed to terminate with a valid estimate.

\section{Statistical Model Checking with Rewards\label{sec:SMC}}

The notions of probability estimation used in standard SMC can be
adapted to estimate the expected reward of a trace. Given a function
$R(\omega)\in[a,b]$, $a,b$ finite, that assigns a total reward to
simulation trace $\omega$, the expected reward may be estimated by
$\frac{1}{N}\sum_{i=1}^{N}R(\omega_{i})$, where $\omega_{1},\dots,\omega_{N}$
are statistically independent simulation traces. Since rewards may
take values outside $[0,1]$, we must use Hoeffding's generalisation
of (\ref{eq:chernoff}) to bound the errors \cite{Hoeffding1963}.
To guarantee $\mathrm{P}(\mid\hat{r}-r\mid\geq\varepsilon)\leq\delta$,
where $r$ and $\hat{r}$ are respectively the true and estimated
values of expected reward, $N$ is required to satisfy the relation

\begin{equation}
N\geq\left\lceil \ln(2/\delta)\times(a-b)^{2}/(2\varepsilon^{2})\right\rceil .\label{eq:hoeffding}
\end{equation}

For non-trivial problems the values of $a$ and $b$ are usually not
known, while guaranteed a priori bounds (e.g., by assuming maximum
or minimum possible rewards on each step) may be too conservative
to be useful. While it is possible to develop a strategy using a posteriori
estimates of $a$ and $b$, i.e., based on $\max_{i\in\{1,\dots,N\}}(R(\omega_{i}))$
and $\min_{i\in\{1,\dots,N\}}(R(\omega_{i}))$, we see that $N$ depends
on the ratio of the absolute error $\varepsilon$ to the range of
values $(a-b)$. The confidence of estimates of rewards may therefore
be specified a priori as a percentage of the maximum range of the
support of $R$. We adopt this idea in Algorithm \ref{alg:smartreward},
where we use (\ref{eq:chernoff}) and (\ref{eq:NMChernoff}) and assume
that $\varepsilon$ expresses a percentage as a fraction of $1$.

\subsection*{Rewards Properties}

The rewards properties commonly used in numerical model checking are
based on an extension of the logic PCTL \cite{KwiatkowskaNormanParker2007}.
This extension defines \emph{instantaneous} rewards (the average reward
assigned to the $k^{\mathrm{th}}$ state of all traces, denoted $\mathbf{I}^{k}$),
\emph{cumulative} rewards (the average total reward accumulated up
to the $k^{\mathrm{th}}$ state of all traces, denoted $\mathbf{C}^{k}$)
and \emph{reachability} rewards (the average accumulated reward of
traces that eventually satisfy property $\varphi$, denoted $\mathbf{F}\varphi$).
Instantaneous and cumulative rewards are based on finite traces and
can be immediately approximated by sampling, using (\ref{eq:chernoff})
and (\ref{eq:NMChernoff}) to bound the errors. Reachability rewards
are based on unbounded $\mathbf{F}$ and require additional consideration.

\smallskip{}

By the definition of reachability rewards \cite{KwiatkowskaNormanParker2007},
properties that are not satisfied with probability $1$ are assigned
infinite reward. The rationale behind this is that if $\mathrm{P}(\mathbf{F}\varphi)<1$,
there must exist an infinite path that does not satisfy $\varphi$,
whose rewards will accumulate infinitely. This definition is somewhat
arbitrary, since rewards are not constrained to be positive---an infinite
sum of positive and negative values can equate to zero---and it is
also possible for an infinite sum of positive values to converge,
as in the case of discounted rewards.

The definition of reachability rewards makes sense in the context
of numerical model checking, where paths are not considered explicitly
and unbounded properties can be quantified with certainty, but it
causes problems for sampling. In particular, using sampling alone
it is not possible to say with certainty whether $\mathrm{P}(\mathbf{F}\varphi)=1$,
even if every observed trace of finitely many satisfies $\varphi$.
Without additional guarantees, the random variable from which samples
are drawn could include the value infinity, giving it infinite variance.
Statistical error bounds, which generally rely on an underlying assumption
of finite variance, will therefore not be correct without additional
measures.

To accommodate the standard definition of reachability rewards, our
solution is to implement $\mathbf{F}\varphi$ as $\mathbf{F}^{k}\varphi$,
i.e., bounded reachability, with an auxiliary hypothesis test to assert
that $\mathrm{P}(\mathbf{F}\varphi)=1$ is true. A positive result
is thus an estimate within user-specified confidence plus an accepted
hypothesis within other user-specified confidence. A negative result
is a similar estimate, but with an hypothesis that is not accepted.
This approach is consistent with intuition and with the SMC ethos
to provide results within statistical confidence bounds. The hypothesis
test may be implemented in any number of standard ways. Our implementation
uses a convenient normal approximation model, which we describe in
Section \ref{sec:smartest}.

In practice, the bound $k$ for reachability rewards is set much longer
than it is supposed necessary to satisfy $\varphi$ and the hypothesis
is of the form $\mathrm{P}(\mathbf{F}^{k}\varphi)\geq p_{0}$, $p_{0}\lessapprox1$.
Intuitively, the more traces of length $\leq k$ that satisfy $\varphi$,
the more confident we are that $\mathrm{P}(\mathbf{F}\varphi)\geq p_{0}$
is true. Traces that fail to satisfy $\varphi$ after $k$ steps may
nevertheless satisfy $\varphi$ if allowed to continue, hence the
value of $p_{0}$ defines how certain we wish to be after $k$ steps.
If the hypothesis is rejected, we may either conclude that the average
reward is infinite (by definition), accept the calculated average
reward as a lower bound or increase $k$ and try again.

Our SMC engine quits as soon as a property is satisfied or falsified,
so there is very little penalty in setting $k$ large when we require
high confidence, i.e., when $p_{0}\approx1$. Simulations that satisfy
the property will take only as long as necessary, independent of $k$,
while those that do not satisfy the property will be few because the
auxiliary hypothesis is falsified quickly when $p_{0}$ is close to
1.

Finding schedulers that optimise the rewards defined in \cite{KwiatkowskaNormanParker2007}
has a conceptual advantage over finding schedulers that optimise the
probability of a property. This is because the effective probability
of these rewards properties is always close to $1$. In the case of
instantaneous and cumulative rewards, traces are not filtered with
respect to a property, so the probability of acceptance is $1$. In
the case of reachability rewards, either nearly all traces satisfy
the property (`nearly' because the auxiliary hypothesis test allows
for the case that not all traces satisfy the property) or the reward
is assumed to be infinite. Hence, the case of probabilities significantly
less than $1$ does not have to be quantified, just detected. The
consequence of this, according to (\ref{eq:probgoodsched}), is that
the simulation budget to generate the initial candidate set can be
allocated entirely to schedulers, i.e., $N=1$ and $M=N_{\max}$.

\section{Smart Reward Estimation Algorithm\label{sec:smartest}}

Algorithm \ref{alg:smartreward} finds schedulers that maximise rewards.
The algorithm to minimise rewards follows intuitively: replace instances
of `$\max$' with `$\min$' in lines \ref{alg:max2min2}, \ref{alg:max2min3},
\ref{alg:max2min4} and the Output line, and replace line \ref{alg:max2min5}
with $S\leftarrow\{\sigma\in S\mid\sigma=Q'(n)\wedge n\in\{1,\dots,\lceil|S|/2\rceil\}\}$.

The reward property $\rho$ may be of type instantaneous, cumulative
or reachability, which are denoted $\mathbf{I}^{k}\varphi$, $\mathbf{C}^{k}\varphi$
and $\mathbf{F}^{k}\varphi$, respectively, to unify the description.
The reward function $\mathcal{R}_{\rho}:\mathbb{N}\times\Omega\rightarrow\mathbb{Q}$
maps the identifier of a scheduler and a trace to a reward, given
reward property $\rho$. In the case of $\mathbf{I}^{k}\varphi$ and
$\mathbf{C}^{k}\varphi$, $k$ is the standard user-specified parameter
for these rewards and $\varphi$ is implicitly $\mathbf{G}^{k}\mathit{true}$.
In the case of $\mathbf{F}^{k}\varphi$, $\varphi$ is user-specified
and $k$ is set as large as feasible to satisfy the hypothesis $\mathrm{P}(\mathbf{F}^{k}\varphi)\geq p_{0}$,
with confidence defined by $\alpha$ (described below). Given that
our actual requirement is that $\mathrm{P}(\mathbf{F}\varphi)=1$,
both $p_{0}$ and $\alpha$ will typically be close to $1$, such
that very few traces will be necessary to falsify the hypothesis.

\begin{algorithm}[H]
\KwIn{\\\Indp

$\mathcal{M}$: an MDP 

$\rho\in\{\mathbf{I}^{k}\varphi,\mathbf{C}^{k}\varphi,\mathbf{F}^{k}\varphi\}$:
a reward property

$\mathcal{R}_{\rho}$: the reward function for $\rho$

$H_{0}:\mathrm{P}(\mathbf{F}^{k}\varphi)\geq p_{0}$: the auxiliary
hypothesis

$z(\alpha)$: confidence of $H_{0}$, the normal quantile of order
$\alpha$

$\varepsilon,\delta$: the reward estimation Chernoff bound

$N_{\max}>\ln(2/\delta)/(2\varepsilon^{2})$: the per-iteration budget

}\KwOut{\\\Indp

$\hat{r}_{\overline{\max}}\approx r_{\max}$, where $r_{\overline{\max}}\approx r_{\max}$
and $\mathrm{P}(|r_{\overline{\max}}-\hat{r}_{\overline{\max}}|\geq\varepsilon)\leq\delta$\label{alg:max2min1}

}\BlankLine

$N\leftarrow1,\, M\leftarrow N_{\max}$\label{alg:stageiistart}

$S\leftarrow\{M\textnormal{ seeds chosen uniformly at random}\}$

$\forall\sigma\in S,\forall j\in\{1,\dots,N\}:\omega_{j}^{\sigma}\leftarrow\mathrm{Simulate}(\mathcal{M},\varphi,\sigma)$ 

$Q\leftarrow\{(\sigma,q)\mid\sigma\in S\wedge\mathbb{Q}\ni q=\sum_{j=1}^{N}\mathcal{R}_{\rho}(\sigma,\omega_{j}^{\sigma})/N\}$\label{alg:stageiiend}

$\forall\sigma\in S:\;\mathit{trues}(\sigma)\leftarrow0$\label{alg:trues}

$\mathit{samples}\leftarrow0,\,\mathit{conf}\leftarrow1,\, i\leftarrow0$\label{alg:loopinit}

\label{alg:stageiiistart}

\While{$\mathit{conf}>\delta\wedge S\neq\emptyset$}{

$i\leftarrow i+1$

$M_{i}\leftarrow|S|,\, N_{i}\leftarrow0$

\While{$\mathit{conf}>\delta\wedge N_{i}<\lceil N_{\max}/M_{i}\rceil$}{\label{alg:quitiiistart}

$N_{i}\leftarrow N_{i}+1$

$\mathit{conf}\leftarrow1-(1-e^{-2\epsilon^{2}N_{i}})^{M_{i}}$

$\forall\sigma\in S:\;\omega_{N_{i}}^{\sigma}\leftarrow\mathrm{Simulate}(\mathcal{M},\varphi,\sigma)$

$\mathit{samples}\leftarrow\mathit{samples}+1$\label{alg:quitiiiend}

}

$Q\leftarrow\{(\sigma,q)\mid\sigma\in S\wedge\mathbb{Q}\ni q=\sum_{j=1}^{N_{i}}\mathcal{R}_{\rho}(\sigma,\omega_{j}^{\sigma})/N_{i}\}$\label{alg:stageiiisort}

$\sigma_{\max}\leftarrow\mathop{\arg\max}_{\sigma\in S}\, Q(\sigma)$\label{alg:max2min2}

$\hat{r}_{\overline{\max}}\leftarrow Q(\sigma_{\max})$\label{alg:max2min3}

$\forall\sigma\in S,j\in\{1,\dots,N_{i}\}:\;\mathit{trues}(\sigma)=\mathit{trues}(\sigma)+\mathbf{1}(\omega_{j}^{\sigma}\models\varphi)$

$Q':\{1,\dots,|S|\}\rightarrow S$ is an injective function s.t.

\nonl\quad{}$\forall(n,\sigma),(n',\sigma')\in Q':\; n>n'\implies Q(\sigma)\geq Q(\sigma')$

$S\leftarrow\{\sigma\in S\mid\sigma=Q'(n)\wedge n\in\{\lfloor|S|/2\rfloor,\dots,|S|\}\}$\label{alg:max2min5}\label{alg:stageiiiend}

}

$Z\leftarrow(\mathit{trues}(\sigma_{\max})-\mathit{samples}\times p_{0})/\sqrt{\mathit{samples}\times p_{0}\,(1-p_{0})}$\label{alg:max2min4}\label{alg:teststart}

\If{$Z\leq z(\alpha)$}{$H_{0}$ is rejected\label{alg:testend}}

\caption{Reward Estimation\label{alg:smartreward}}
\end{algorithm}

The initial candidate set of schedulers and corresponding estimates
are generated in lines \ref{alg:stageiistart} to \ref{alg:stageiiend}.
Applying (\ref{eq:probgoodsched}), $1$ simulation is performed using
each of $N_{\max}$ schedulers chosen at random. The function $Q$
maps schedulers to their current estimate. A number of initialisations
take place in lines \ref{alg:trues} to \ref{alg:loopinit}.

The function $\mathit{trues}$ is used by the auxiliary hypothesis
test and counts the total number of traces per scheduler that satisfy
the property. The variable $\mathit{samples}$ is also used by the
auxiliary hypothesis test and counts the total number of traces used
by any scheduler. The value of $\mathit{conf}$, initialised to $1$
to ensure at least one iteration, is the probability that the estimates
exceed their specified bounds (defined by $\varepsilon$), given the
current number of simulations. The main loop (lines \ref{alg:stageiiistart}
to \ref{alg:stageiiiend}) terminates when $\mathit{conf}$ is less
than or equal to the specified probability $\delta$. Typically, the
per-iteration budget will be such that the required confidence is
reached according to (\ref{eq:NMChernoff}) before the candidate set
is reduced to a single element. Lines \ref{alg:quitiiistart} to \ref{alg:quitiiiend}
contain the main simulation loop, which quits as soon as the required
confidence is reached. Lines \ref{alg:stageiiisort} to \ref{alg:stageiiiend}
order the results by estimated reward and select the upper quantile
of schedulers.

The auxiliary hypothesis test necessary for reachability rewards is
provided in lines \ref{alg:teststart} to \ref{alg:testend}. To test
$\mathrm{P}(\mathbf{F}^{k}\varphi)\geq p_{0}$, it considers the error
statistic $Z=\mathit{samples}\times(\hat{p}_{\varphi}-p_{0})/\sqrt{\mathit{samples}\times p_{0}(1-p_{0})}$,
where $\hat{p}_{\varphi}=\mathit{trues}(\sigma_{\max})/\mathit{samples}$
is the estimate of $\mathrm{P}(\mathbf{F}^{k}\varphi)$. For typical
values of $\mathit{samples}$, the distribution of $Z$ is well approximated
by a normal with mean $=0$ and variance $=1$ when the expectation
of $\hat{p}_{\varphi}$, denoted $\mathrm{E}(\hat{p}_{\varphi})$,
is equal to $p_{0}$. To test the hypothesis with confidence $\alpha$,
the algorithm compares the statistic $Z$ with the standard normal
quantile of order $\alpha$, denoted $z(\alpha)$. The value of $z(\alpha)$
may be drawn from a table or approximated numerically. If $\mathrm{E}(\hat{p}_{\varphi})\geq p_{0}$,
the value of $Z$ will be $\geq z(\alpha)$ with probability $\geq\alpha$.

To simplify the presentation of our algorithm, the auxiliary hypothesis
test is also used by the instantaneous and cumulative rewards. In
these latter cases, however, the test is guaranteed to be satisfied.
Also for the sake of simplicity, the algorithm assumes that all simulation
traces reach the $k^{\mathrm{th}}$ state without halting.

\section{Case Studies\label{sec:experiments}}

We have implemented Algorithms \ref{alg:simulate} and \ref{alg:smartreward}
in our statistical model checking platform, \textsc{Plasma}$^{\ref{fn:plasma}}$,
and take advantage of its distributed verification algorithm on the
\textsc{Igrida} parallel computational grid%
\footnote{igrida.gforge.inria.fr%
}. All timings are based on 64 simulation cores. The following results
demonstrate typical performance on a selection of standard case studies,
including one with intractable state space. We necessarily use models
whose expected rewards can be calculated or inferred using numerical
methods, but observe that this does not give our algorithms any advantage.
The models and properties can be found on our website$^{\ref{fn:plasma}}$
and are illustrated in detail on the \textsc{Prism} case studies website%
\footnote{www.prismmodelchecker.org/casestudies\label{fn:prism}%
}.

In most instances we were able to achieve accurate results with a
relatively modest per-iteration simulation budget of $N_{\max}=10^{5}$
simulations, using a Chernoff bound of $\varepsilon=\delta=0.01$.
In the case of the gossip protocol (Section \ref{sec:gossip}), this
budget was apparently not sufficient for all considered parameters.
We nevertheless claim that the results provide useful conservative
bounds. Note that for all reachability rewards we made the value of
$k$ in the auxiliary hypothesis test sufficiently large to ensure
that all traces satisfied the property, giving us maximum confidence
for the specified budget.

\subsection{Network Virus Infection\label{sec:virus}}

\begin{figure}
\begin{minipage}[t]{0.48\columnwidth}%
\begin{center}
\includegraphics[width=0.9\columnwidth]{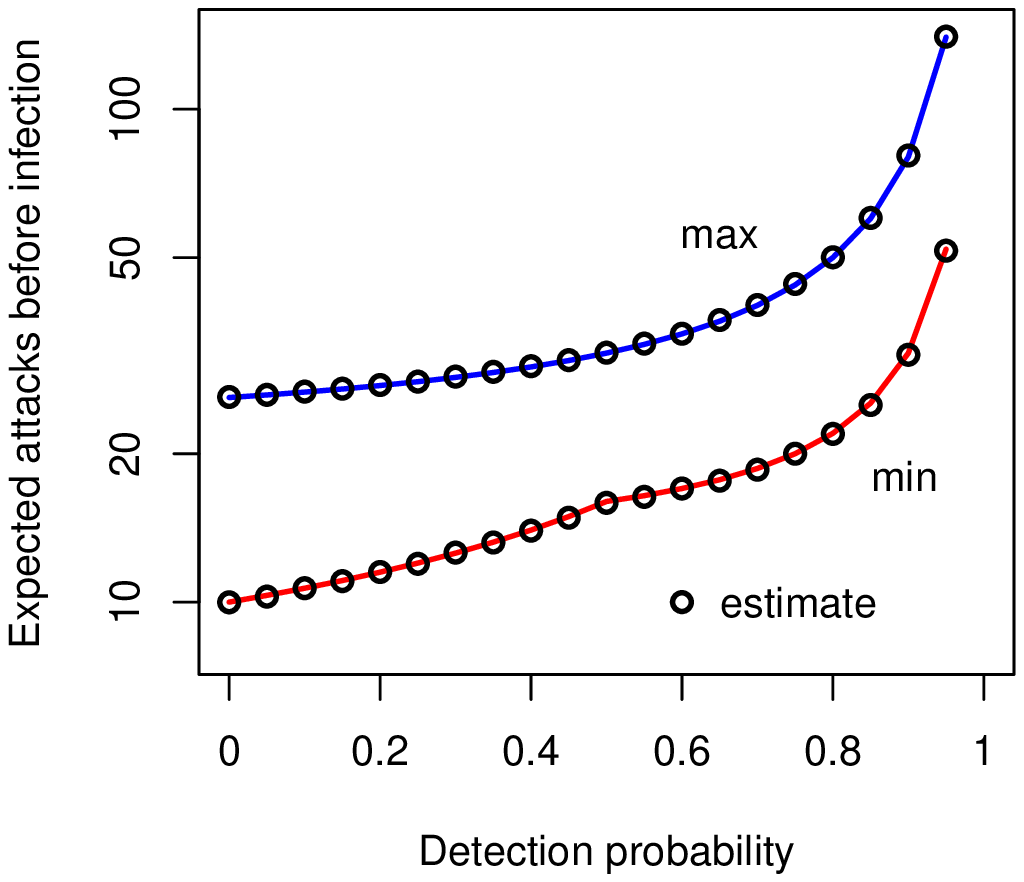}\caption{Network virus infection.\label{fig:virus}}

\par\end{center}%
\end{minipage}\quad{}%
\begin{minipage}[t]{0.48\columnwidth}%
\begin{center}
\includegraphics[width=0.9\columnwidth]{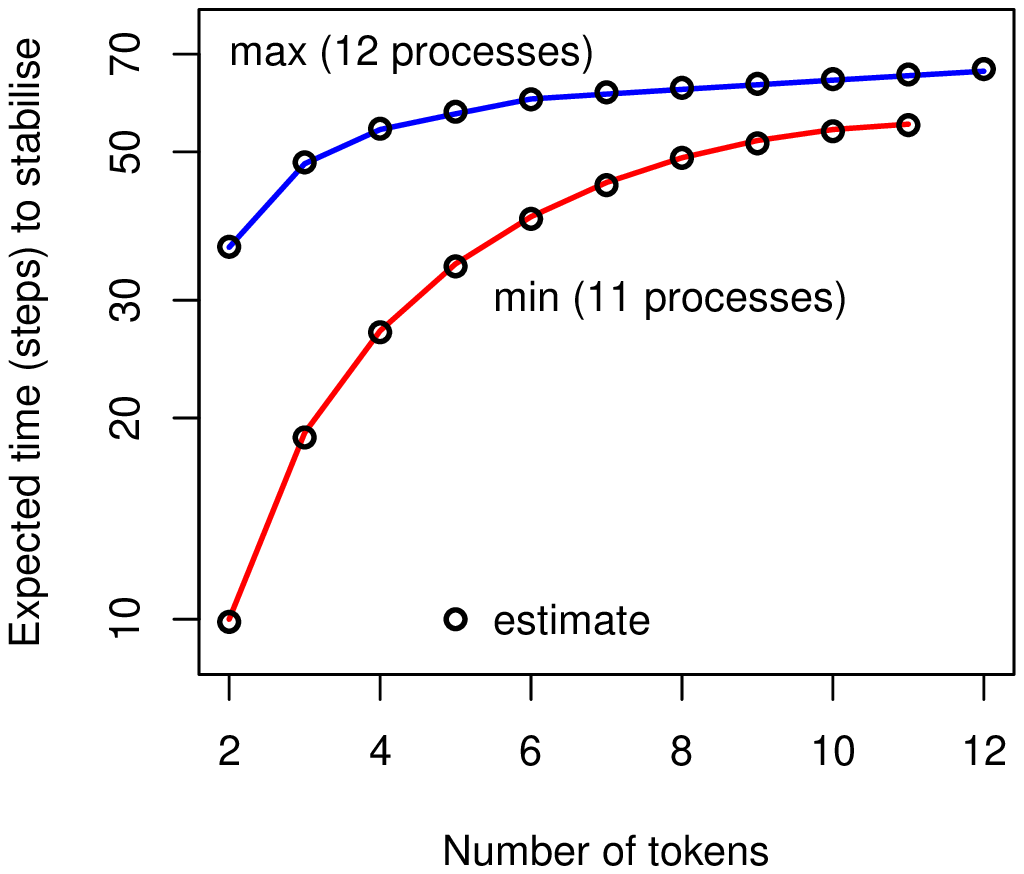}\caption{Self-stabilising protocol.\label{fig:selfstab}}

\par\end{center}%
\end{minipage}
\end{figure}

Our network virus infection case study is based on \cite{KwiatkowskaNormanParkerVigliotti2009}
and initially comprises the following sets of linked nodes: a set
containing one node infected by a virus, a set with no infected nodes
and a set of uninfected barrier nodes which divides the first two
sets. A virus chooses which node to infect nondeterministically. A
node detects a virus probabilistically and we vary this probability
as a parameter for barrier nodes. Figure \ref{fig:virus} illustrates
the results of using a reachability reward property to estimate the
maximum and minimum expected number of detected attacks before a particular
node is infected. Each point required approximately 15 seconds of
simulation time. The solid lines indicate values calculated numerically.
All estimates are within $\pm1\%$ of the true values.

\subsection{Self Stabilisation\label{sec:self-stab}}

The self-stabilising protocol of \cite{IsraeliJalfon1990} works asynchronously
to ensure that a number of networked processes share a single `privileged
status' token fairly. The protocol is designed to reach this dynamical
state even if initially there are several tokens in the ring. For
various numbers of processes, we used reachability properties to estimate
the maximum and minimum expected number of steps to reach stability,
given different initial numbers of tokens. Figure \ref{fig:selfstab}
plots typical results: the maximum values for 12 processes and the
minimum values for 11 processes. Individual estimates required between
1 and 3 minutes of simulation time. The solid lines indicate values
calculated numerically. All estimates are within $\pm1\%$ of the
true values.

\subsection{Gossip Protocol\label{sec:gossip}}

\begin{wrapfigure}{o}{0.5\textwidth}%
\includegraphics[width=0.9\linewidth]{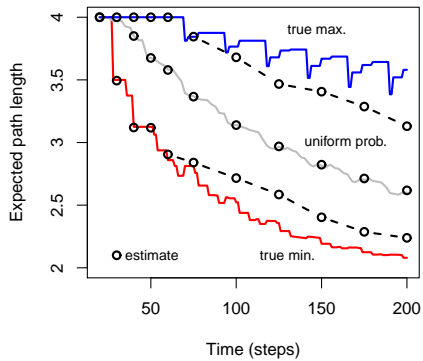}

\caption{Gossip protocol.\label{fig:gossip}}
\end{wrapfigure}%

The gossip protocol of \cite{KwiatkowskaNormanParker2008} uses local
connectivity to propagate information globally. Using a reachability
reward property, our algorithms accurately estimate the expected minimum
and maximum number of rounds necessary for the network to become connected
as 1.486 and 4.5, compared to correct values 1.5 and 4.5. The average
simulation time per estimate was approximately 1 minute.

Figure \ref{fig:gossip} plots the maximum and minimum estimated path
length at different time steps, using an instantaneous reward property.
The figure also plots the average estimates of the initial candidate
set (generated in lines \ref{alg:stageiistart} to \ref{alg:stageiiend}
of Algorithm \ref{alg:smartreward}). This average corresponds to
the expected reward using the uniform probabilistic scheduler. The
solid lines give the values calculated using numerical algorithms.
The true value for the uniform probabilistic scheduler is calculated
numerically by treating the MDP as a DTMC. We see that the estimates
of maximum and minimum expected reward are accurate up to about 75
time steps, but less so above this value. The estimates are nevertheless
guaranteed by (\ref{eq:NMChernoff}) not to exceed the true values
by more than a factor of $1+\varepsilon$ with probability $\delta$.
Finally, we note that the average estimate of the initial candidate
set is consistently accurate, demonstrating that Algorithm \ref{alg:simulate}
is sampling correctly from scheduler space.

\subsection{Choice Coordination\label{sec:choice}}

To demonstrate the scalability of our approach, we consider instances
of the choice coordination model of \cite{NdukwuMcIver2010} with
$\mathit{BOUND}=100$. This value makes most of the models intractable
to numerical model checking, however it is possible to infer the correct
values of rewards from tractable instances. The chosen reachability
property gives the expected minimum number of rounds necessary for
a group of tourists to meet. The following table summarises the results:

\begin{center}
\begin{tabular}{|c|c|c|c|c|c|c|c|c|c|}
\hline 
Number of tourists & 2 & 3 & 4 & 5 & 6 & 7 & 8 & 9 & 10\tabularnewline
\hline 
Minimum number of rounds to converge & 4.0 & 5.0 & 7.0 & 8.0 & 10.0 & 11.0 & 12.0 & 13.0 & 14.0\tabularnewline
\hline 
\end{tabular}
\par\end{center}

All the estimates are exactly correct, while the average time to generate
each result was just 8 seconds.

\section{Prospects and Challenges \label{sec:prospects}}

In this work we have focused on estimating the expected value of optimal
rewards. We believe the same techniques may be immediately extended
to sequential hypothesis testing, as in \cite{LegaySedwardsTraonouez2014}
and \cite{D'ArgenioLegaySedwardsTraonouez2015}. Ongoing work suggests
that estimating rewards in continuous time models will also be feasible.

Overall, our case studies demonstrate that our approach is effective
and can be efficient with state space that is intractable to numerical
methods. While we do not yet provide confidence with respect to optimality,
our techniques nevertheless generate useful conservative bounds with
correct statistical guarantees of accuracy: the estimate will be greater
(less) than the true maximum (minimum) expected reward by a factor
of $\geq1+\varepsilon$ with probability $\leq\delta$.

Figure \ref{fig:gossip} illustrates circumstances where the chosen
per-iteration budget of $10^{5}$ is apparently not sufficient to
explore the tails of the distribution of schedulers. Merely increasing
the budget will not in general be adequate to address this problem,
since near-optimal schedulers may be arbitrarily rare. Our proposed
solutions are to (\emph{i}) construct composite schedulers and (\emph{ii})
sample from a property-specific subset of schedulers.

\section*{Acknowledgement}

This work was partially supported by the European Union Seventh Framework
Programme under grant agreement no. 295261 (MEALS).

\bibliographystyle{abbrv}
\bibliography{MDP_rewards}

\end{document}